\documentclass[12pt]{article} 
\usepackage{epsfig}
\newcommand{\eq}{\begin{equation}} 
\newcommand{\eqx}{\end{equation}} 
\newcommand{\eqn}{\begin{eqnarray}} 
\newcommand{\eqnx}{\end{eqnarray}} 
 
\newcommand{\fsn}{$g_1(x,Q^2)\,\,\,\,$} 
\newcommand{\nnu}{\nu^{\prime}}

\newcommand{\nin}{\noindent} 
\def\lapproxeq{\lower .7ex\hbox{$\;\stackrel{\textstyle
<}{\sim}\;$}}
\def\gapproxeq{\lower .7ex\hbox{$\;\stackrel{\textstyle
>}{\sim}\;$}}
\begin{document} 
\nin 


%
%
\mbox{}
\\

\vskip2cm
\begin{center} 
{\Large \bf Spin structure function $g_1(x,Q^2)$  and
the DHGHY integral $I(Q^2)$ at low~$Q^2$:
predictions from the GVMD model
}\\ 
\vspace{10mm} 
 
{\Large Barbara~Bade\l{}ek $^{\ast,\,\dag}$}, 
{\Large Jan~Kwieci\'nski $^{\ddag}$} and 
{\Large Beata~Ziaja $^{\ddag,\,\S,\,\$,\,}$} 
\footnote {e-mail addresses: badelek@fuw.edu.pl,~~jan.kwiecinski@ifj.edu.pl,
~~ziaja@tsl.uu.se}\\ 
\end{center} 

{\footnotesize 

\vspace{3mm} 
\noindent 
           $^{\ast}$ \it Institute of Experimental Physics, Warsaw University,
	               Ho\.za 69, 00-681 Warsaw, Poland\\  
           $^{\dag}$ \it Department of Physics, Uppsala University, 
	               P.O. Box 530, S-751 21 Uppsala, Sweden \\ 
           $^{\ddag}$ \it Department of Theoretical Physics, 
	               H. Niewodnicza\'nski Institute of Nuclear Physics, \\ 
                 \it \mbox{}\hskip3mm Radzikowskiego 152,  31-342 Cracow, Poland\\ 
           $^{\S}$ \it Department of Biochemistry, BMC, Uppsala University,
                       Box 576, S-751 23 Uppsala, Sweden\\ 
           $^{\$}$ \it High Energy Physics, Uppsala University, 
	               P.O. Box 535, S-751 21 Uppsala, Sweden\\ 
} 
 
\vspace{5mm} 
\nin 
{\bf Abstract:} 
{\footnotesize  
Theoretical predictions for polarized nucleon structure function $g_1(x,Q^2)$ 
at low $Q^2$ are obtained in the framework of the Generalized 
Vector Meson Dominance model. Contributions from both light and heavy 
vector mesons are evaluated. In the 
photoproduction limit the first moment of $g_1$ 
is related to the static properties of nucleon via the 
Drell-Hearn-Gerasimov-Hosoda-Yamamoto sum rule. 
This property is employed to fix the magnitude of the light vector meson
contribution to $g_1$, using the recent measurements in the region
of baryonic resonances. Results are compared to the data on $g_1(x,Q^2)$.
 Finally, the DHGHY moment function $I(Q^2)$ is calculated, and our theoretical
predictions are confronted with the recent preliminary data obtained at the 
Jefferson Laboratory.
} 
\vspace{6mm} 
\section{Introduction} 

\noindent 
Data on polarized nucleon structure function $g_1(x,Q^2)$ 
are now available in the
region of low values of (negative) four-momentum transfer, 
$Q^2$, \cite{nne143,smc98,ssmc98,hermes}. 
This region is of particular interest since 
nonperturbative mechanisms dominate the particle dynamics there and thus
a transition from soft- to hard physics may be studied.
In the fixed target experiments the low values 
of $Q^2$ are reached simultaneously with the low values of the Bjorken variable, 
$x$, \cite{smc98,ssmc98}, and therefore predictions for 
spin structure functions in both the low $x$ 
and low $Q^2$ region are needed. Partonic contribution to $g_1$ which controls 
the structure function in the deep inelastic domain 
has thus to be suitably extended to the
 low $Q^2$ region. In the  $Q^2$=0 limit, $g_1$ should be a finite function of
$W^2$, free from any kinematic singularities or zeroes.

\noindent
In the previous attempt, \cite{bkk}, $g_1$ at low $x$ and low $Q^2$ was described
within a formalism based on the unintegrated spin dependent parton 
distributions,
incorporating the leading order Altarelli--Parisi evolution and the double
ln$^2$(1/$x$) resummation at low $x$. The ln$^2$(1/$x$) effects are not yet 
significant in the kinematic range of the fixed target ex\-pe\-ri\-ments but the
formalism based on unintegrated parton distributions is very suitable for 
extrapolating $g_1$ to the region of low $Q^2$. Also a VMD-type nonperturbative 
part of $g_1$ was included, its unknown normalisation to be extracted
from the data. The model reproduced general trends in the measurements 
but their statistical accuracy was too low to constrain the VMD contribution. 
However the nonzero and negative value of the VMD contribution was clearly 
preferred. 

\noindent
A convenient way of describing structure functions both in the nonperturbative
and in the scaling (``asymptotic'') region is to employ the
Generalized Vector Meson Dominance (GVMD) model with an infinite number
of vector mesons which couple to a virtual photon (cf.\cite{bauer}). 
The heavy meson ($M_V>Q_0$) contribution is 
directly related to the structure function in the scaling region, $g_1^{AS}$,
described by the QCD improved parton model, suitably extrapolated to the 
low $Q^2$ region. The contribution 
of light ($M_V<Q_0$) vector mesons describes nonperturbative effects and
 vanishes as 1/$Q^4$ for large $Q^2$. At low $Q^2$ these effects are large 
and predominant. Here  $M_V$ denotes the mass of a vector meson. 
The GVMD model was successfully applied to describe the low $Q^2$ behaviour 
of the unpolarized structure function $F_2(x,Q^2)$ \cite{el89,el92}. 

\noindent
In this paper we apply GVMD to evaluate the nonperturbative contributions
to the polarized structure function $g_1(x,Q^2)$ at low values of $Q^2$. 
We start with the formulation of the GVMD representation
of $g_1(x,Q^2)$ and then specify the representation of the light 
and of the heavy vector meson components: Vector Meson Dominance (VMD) for 
the former and asymptotic for the latter one (Section 2). Two different 
parametrizations are used to describe the VMD part of $g_1(x,Q^2)$. 
The asymptotic part is parametrized using the GRSV
fit \cite{grsv2000}. 
Then the 
Drell-Hearn-Gerasimov-Hosoda-Yamamoto (DHGHY) sum rule \cite{DH,GER,HY}
together with measurements in the resonance region are employed
to fix the magnitude of the light vector meson contribution 
to $g_1$ (Section 3). Numerical results
 are discussed and compared to other analyses in Section 4. Conclusions 
and outlook are given in Section 5.
\section{The GVMD representation of the structure fun\-ction $g_1(x,Q^2)$
and the DHGHY sum rule}
In the GVMD model, the spin dependent nucleon structure function $g_1$ 
has the following representation, valid for fixed $W^2\gg Q^2$, 
i.e. small values of $x$, $x=Q^2/(Q^2+W^2-M^2)$: 

\eq
g_1(x,Q^2)=g_1^{L}(x,Q^2)+g_1^{H}(x,Q^2).
\label{gvmd}
\eqx
The first term:
\begin{equation}
g_1^{L}(x,Q^2)=\frac{M\nu}{4\pi}\,\sum_V\,\frac{M^4_V \Delta \sigma_V(W^2)}
{\gamma_V^2(Q^2+M^2_V)^2},
\label{gl}
\end{equation}
%

\noindent
sums up contributions from light vector mesons, $M_V < Q_0$ where  $Q_0^2
\sim$ 1 GeV$^2$ \cite{el92}.
 Here $W$ is the invariant mass of the electroproduced
hadronic system, $\nu=Q^2/2Mx$, and $M$ denotes the nucleon mass.
The constants $\gamma_V^2$ are determined from the leptonic widths of the
vector mesons and the cross sections $\Delta \sigma_V(W^2)$ are combinations
of the total cross sections for the scattering of polarised mesons and nucleons.
They are not known and have to be
parametrized. Following Ref.\ \cite{bkk}, we assume that they can be expressed
through the combinations of nonperturbative parton distributions,
$\Delta p_j^{(0)}(x)$, evaluated at fixed $Q_0^2$~:
{\footnotesize
\eqn
\frac{M\nu}{4\pi}\sum_{V=\rho,\omega}\,\frac{M^4_V \Delta \sigma_V(W^2)}
{\gamma_V^2(Q^2+M^2_V)^2}&=&
C\left[ \frac{4}{9}(\Delta u_{val}^{(0)}(x)+\Delta \bar u^{(0)}(x))
+       \frac{1}{9}(\Delta d_{val}^{(0)}(x)+\Delta \bar d^{(0)}(x))\right]
\frac{M^4_{\rho}}
{(Q^2+M^2_{\rho})^2},\nonumber\\
\frac{M\nu}{4\pi}\,\,\frac{M^4_{\phi} \Delta \sigma_{\phi}(W^2)}
{\gamma_{\phi}^2(Q^2+M^2_{\phi})^2}&=&
C\left[ \frac{1}{9}(2\Delta \bar s^{(0)}(x))\right]\frac{M^4_{\phi}}
{(Q^2+M^2_{\phi})^2}.\label{vmd2}
\eqnx
}

\noindent
It should be noted that the magnitude of the VMD contribution to $g_1(x,Q^2)$
in (\ref{gvmd}) is weighted by an unknown constant $C$.

\noindent
The second term in (\ref{gvmd}), $g_1^{H}(x,Q^2)$, which represents 
the contribution of heavy ( $M_V > Q_0$) vector mesons to $g_1(x,Q^2)$ 
can also be treated as an extrapolation of the QCD improved parton model 
structure function, $g_1^{AS}(x,Q^{2})$, to arbitrary values of $Q^2$.
\noindent
We shall  use a  simplified representation of $g_1^{H}(x,Q^2)$,
\eq
g_1^{H}(x,Q^2)= g_1^{AS}({\bar x},Q^{2}+Q_0^2),
\label{ghu}
\eqx
as it was done in Ref.\ \cite{el92} for the unpolarized structure function
$F_2$. The scaling variable $x$ on the right hand side of Eq.(\ref{ghu})
is replaced by
${\bar x}=(Q^2+Q_0^2)/(Q^2+Q_0^2+W^2-M^2)$.
It follows from equation (\ref{ghu}) that
 $g_1^{H}(x,Q^2)\rightarrow g_1^{AS}(x,Q^{2})$ as $Q^2$ is large.
Substituting $g_1^{H}(x,Q^2)$ in (\ref{gvmd}) with (\ref{ghu})  and
$g_1^{L}(x,Q^2)$ with a sum of (\ref{vmd2}) we get
\eqn
g_1(x,Q^2)&=&g_1^{L}(x,Q^2)+ g_1^{AS}({\bar x},Q^{2}+Q_0^2)=\nonumber \\
&=&C\left[ \frac{4}{9}(\Delta u_{val}^{(0)}(x)+\Delta \bar u^{(0)}(x))
+\frac{1}{9}(\Delta d_{val}^{(0)}(x)+\Delta \bar d^{(0)}(x))\right]
\frac{M^4_{\rho}}{(Q^2+M^2_{\rho})^2}\nonumber\\
&+&C\left[ \frac{1}{9}(2\Delta \bar s^{(0)}(x))\right]\frac{M^4_{\phi}}
{(Q^2+M^2_{\phi})^2}\nonumber \\
&+&g_1^{AS}({\bar x},Q^{2}+Q_0^2).\label{gvmdu}
\eqnx

\noindent
The only free parameter in (\ref{gvmdu}) is the constant $C$. Its value 
may be fixed in the photoproduction limit where the first moment of the 
structure function $g_1(x,Q^2)$ is related to static properties of the
nucleon via the DHGHY sum rule, cf.\ \cite{ioffe,ioffe2}. 
These static properties are the anomalous magnetic moments. 
For a photon frequency $\nu$ defined as $\nu=pq/M$ ($p$ and $q$ are 
energy-momentum four-vectors of the target nucleon and virtual photon 
respectively, and $q^2=-Q^2$), the photon--nucleon scattering 
amplitude, $S_1(\nu,q^2)$, fulfills the dispersion relation
\cite{ioffe}~: 
\eq 
S_1(\nu,q^2)=4\,\int_{-q^2/2M}^{\infty}\nnu d \nnu  
\frac{G_1(\nnu,q^2)}{(\nnu)^2-\nu^2}, 
\label{disp} 
\eqx 
where $G_1(\nu,q^2)$ is a polarized nucleon structure function 
which in the Bjorken limit ($Q^2$, $\nu\rightarrow \infty$ at fixed 
$x$) is related to \fsn as~: 
\eq 
G_1(\nu,q^2)=\frac{M}{\nu}\,g_1(x,Q^2). 
\label{disp2} 
\eqx 
%
As a result of F.E. Low's theorem \cite{low}:
\eq 
S_1(0,0)=-\kappa^2_{p(n)},
\label{s00} 
\eqx 
function $G_1$ fulfills the DHGHY sum rule in the photoproduction limit, 
$Q^2\rightarrow$0:
\eq
\int_0^\infty \,\frac{d\nu}{\nu}\,G_1(\nu ,0) = -{1\over 4}\kappa^2_{p(n)}.
\eqx

\noindent
Here $\kappa_{p(n)}$ is the anomalous magnetic moments of proton (or neutron).
In the limit $\nu\rightarrow 0$ the right hand side of equation (\ref{disp})
transforms to~:
\eq
S_1(0,q^2)=4M\,\int_{Q^2/2M}^{\infty}\,\frac{d\nu}{\nu^2}
\,g_1\left(x(\nu),Q^2\right).
\label{s10}
\eqx
%
Representation (\ref{s10}) of the scattering amplitude
$S_1(0,q^2)$
%
is valid down to certain threshold value of $W_t$, $W_t\gapproxeq 2$ GeV. 
Below $W_t$ the scattering is dominated by baryonic resonances. We have to 
separate these two regions in Eq. (\ref{s10}), cf. \cite{ioffe,ioffe2}. 
The requirement 
$W > W_t$ gives the lower limit for integration over $\nu$ in (\ref{s10}): 
$\nu>\nu_t(Q^2)$, where $\nu_t(Q^2)=(W_t^2+Q^2-M^2)/2M$. With the DHGHY moment 
defined as $I(Q^2)=S_1(0,q^2)/4$, relation (\ref{s10}) may be rewritten as:
\eq
I(Q^2)=I_{res}(Q^2)
+ M\,\int_{\nu_t(Q^2)}^{\infty}\,\frac{d\nu}{\nu^2}\,g_1\left(x(\nu),Q^2\right),
\label{iq2}
\eqx
and the DHGHY sum rule implies that~:
\eq
I(0)=I_{res}(0)
+ M\,\int_{\nu_t(0)}^{\infty}\,\frac{d\nu}{\nu^2}\,g_1\left(x(\nu),0\right)
=-\kappa^2_{p(n)}/4.
\label{i0}
\eqx

\noindent
Here $I_{res}(Q^2)$ denotes the contribution from resonances.
Substituting $g_1\left(x(\nu),0\right)$ in Eq. (\ref{i0}) by Eq. (\ref{gvmdu}) 
at $Q^2=0$ and performing the integral in (\ref{i0}), we may 
obtain the value of the constant $C$ if $I_{res}(0)$ is known e.g. from 
measurements.

%
%
%
%
\noindent
\begin{figure}[t]
\begin{center}
\epsfig{width=13cm, file=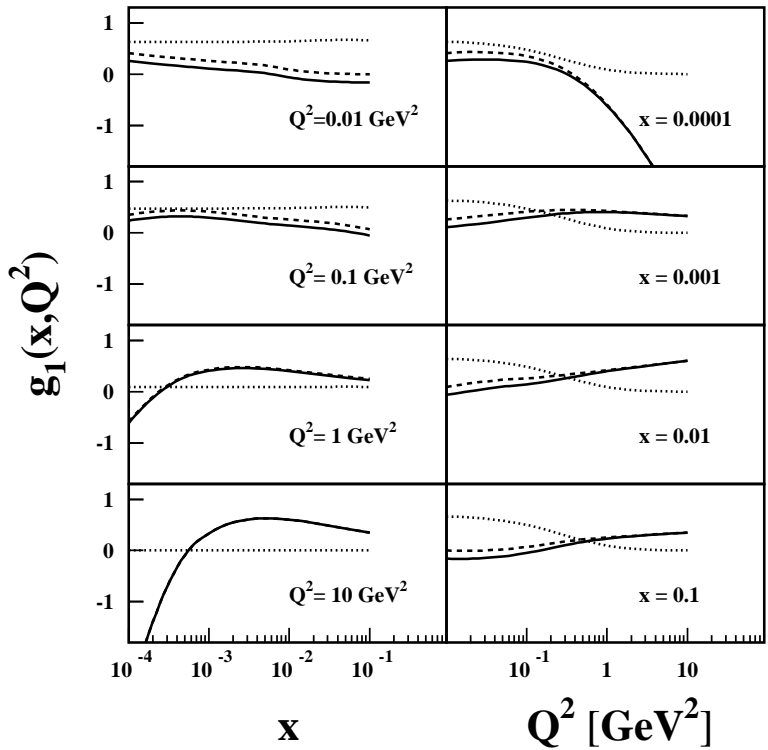}\\
\end{center}
\vskip-1cm
\caption{Values of $g_1$ for the proton as a function of $x$ and $Q^2$.
The asymptotic contribution, $g_1^{AS}$, is marked with broken lines,
the VMD part, $g_1^{L}$, with dotted lines and the continuous curves mark 
their sum, according to eq. (\ref{gvmdu}). 
}
\vskip1cm
\label{fig0}
\end{figure}
\noindent
\begin{figure}[t]
\begin{center}
\epsfig{width=13cm, file=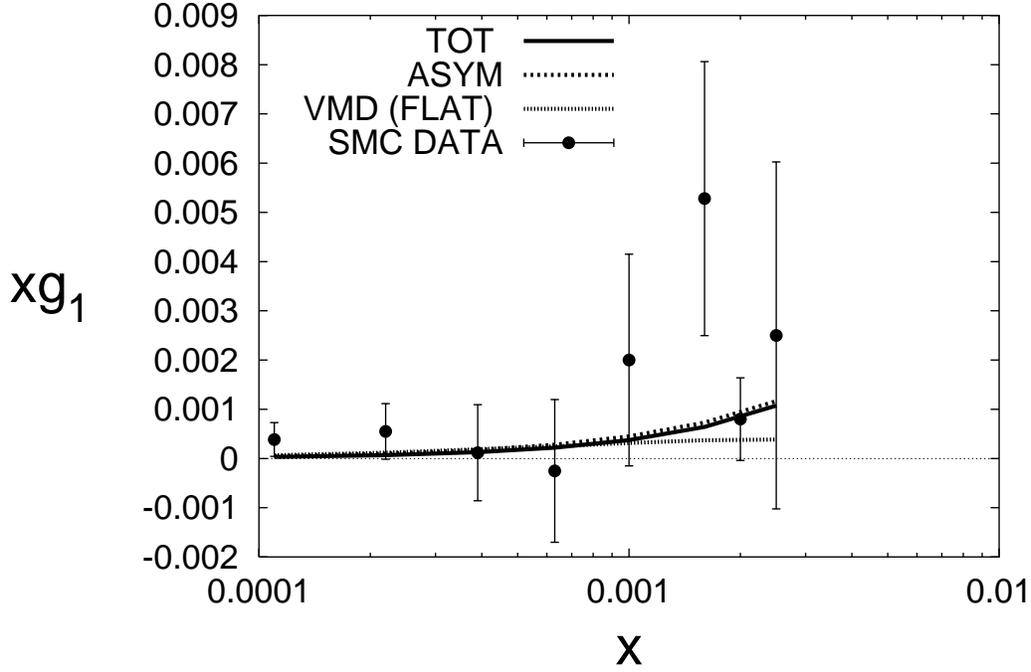}\\
\vskip1cm
\epsfig{width=13cm, file=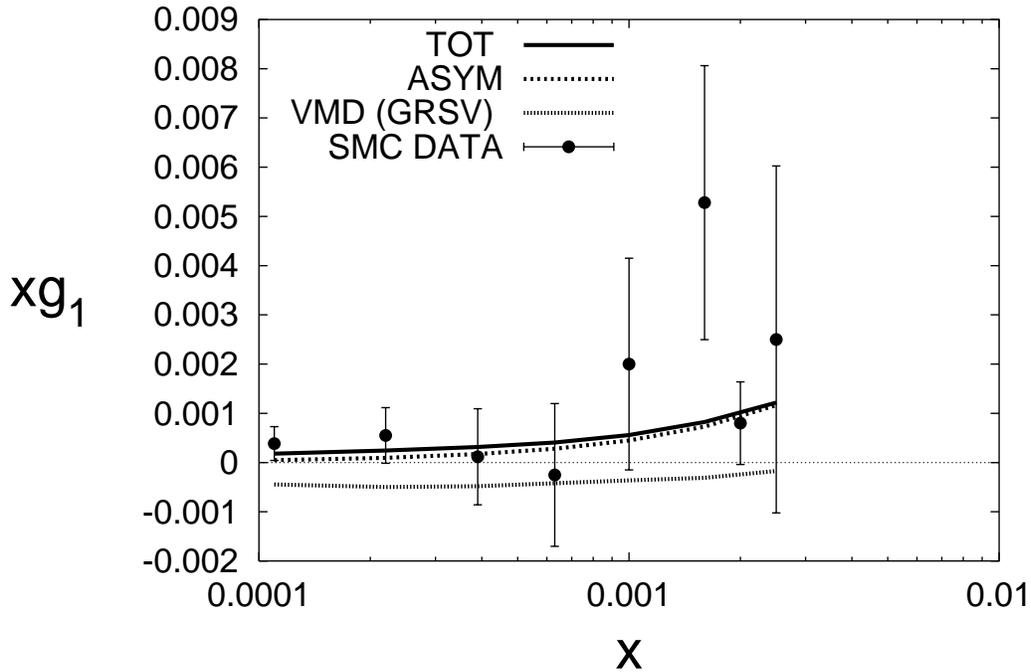}\\
\end{center}
\caption{Values of $xg_1$ for the proton as a function of $x$ at the
measured values of $Q^2$ in the non-resonant region, $x<x_t=Q^2/2M\nu_t(Q^2)$.
Upper plot corresponds to the VMD part parametrized using Eq.(\ref{dpi0}),
lower plot corresponds to the GRSV parametrization \cite{grsv2000}
of the VMD input. 
The $g_1^{AS}$ in both plots has been
calculated using the GRSV fit for standard scenario at the 
NLO accuracy. Contributions of the VMD and of the $xg_1^{AS}$
are shown separately. Points are the SMC measurements at 
$Q^2 <$ 1 GeV$^2$, \cite{ssmc98}; errors are total. The
curves have been calculated at the measured $x$ and $Q^2$ values. }
\label{fig1}
\end{figure}
\noindent
\begin{figure}[t]
\begin{center}
\epsfig{width=13cm, file=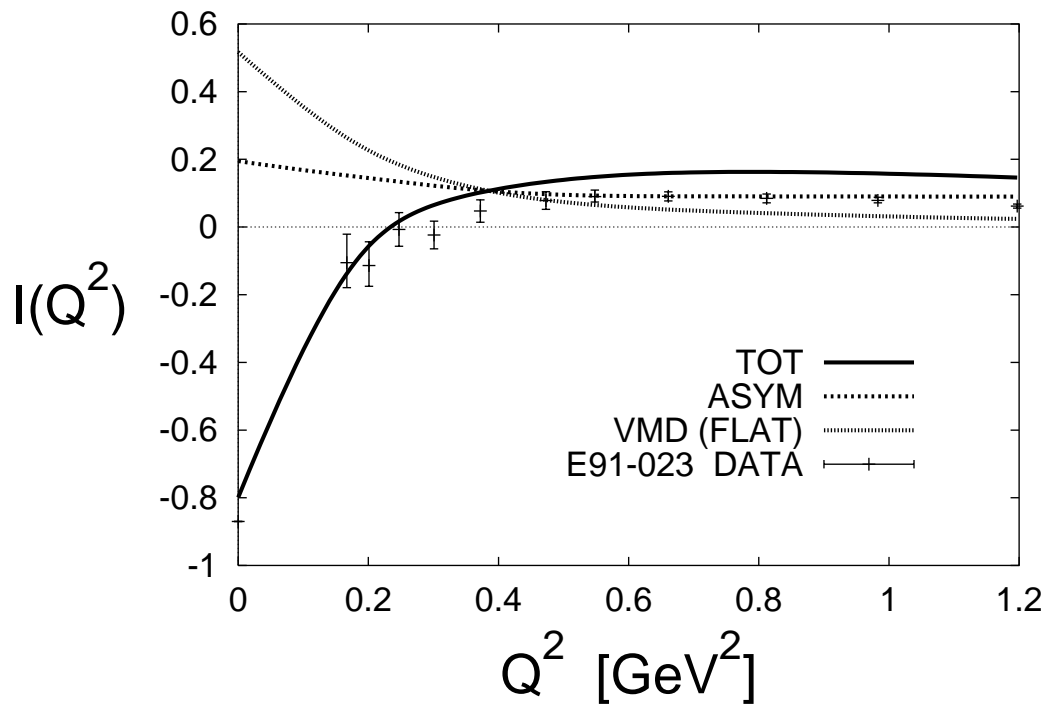}\\
\vskip1cm
\epsfig{width=13cm, file=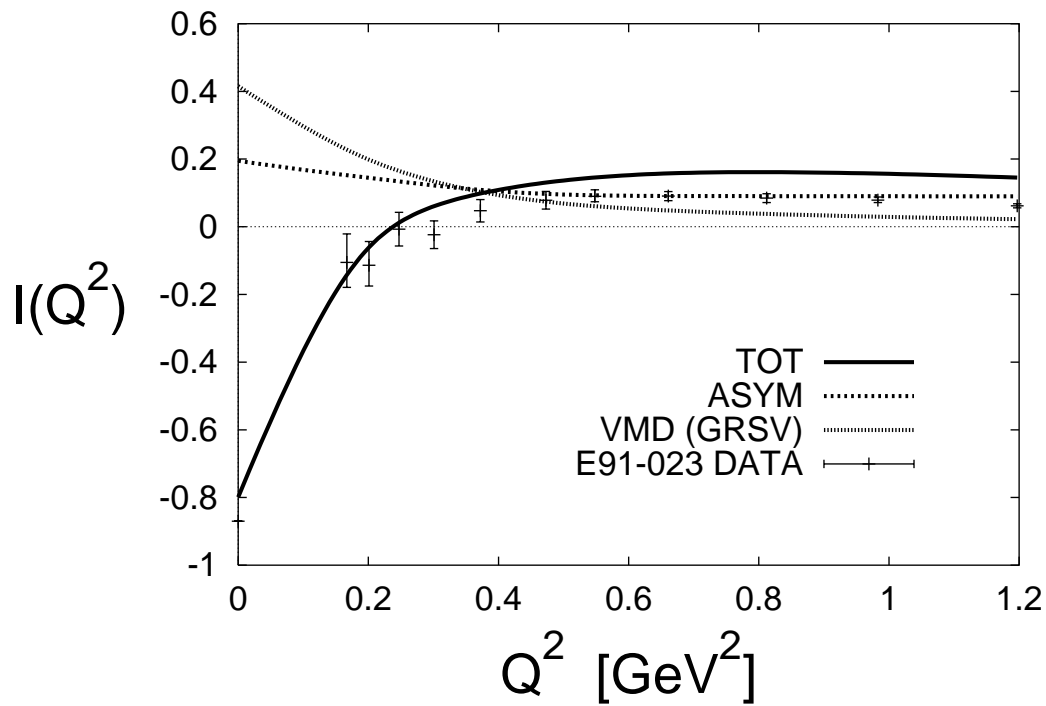}\\
\end{center}
\caption{The DHGHY moment $I(Q^2)$ for the proton. Details concerning
the curves are as in Fig.\ref{fig1}. Points give the contribution of
resonances as measured by the JLAB E91-023 
experiment \cite{e91-023} 
at $W < W_t(Q^2)$.}
\label{fig2}
\end{figure}
\noindent
\begin{figure}[t]
\begin{center}
\epsfig{width=13cm, file=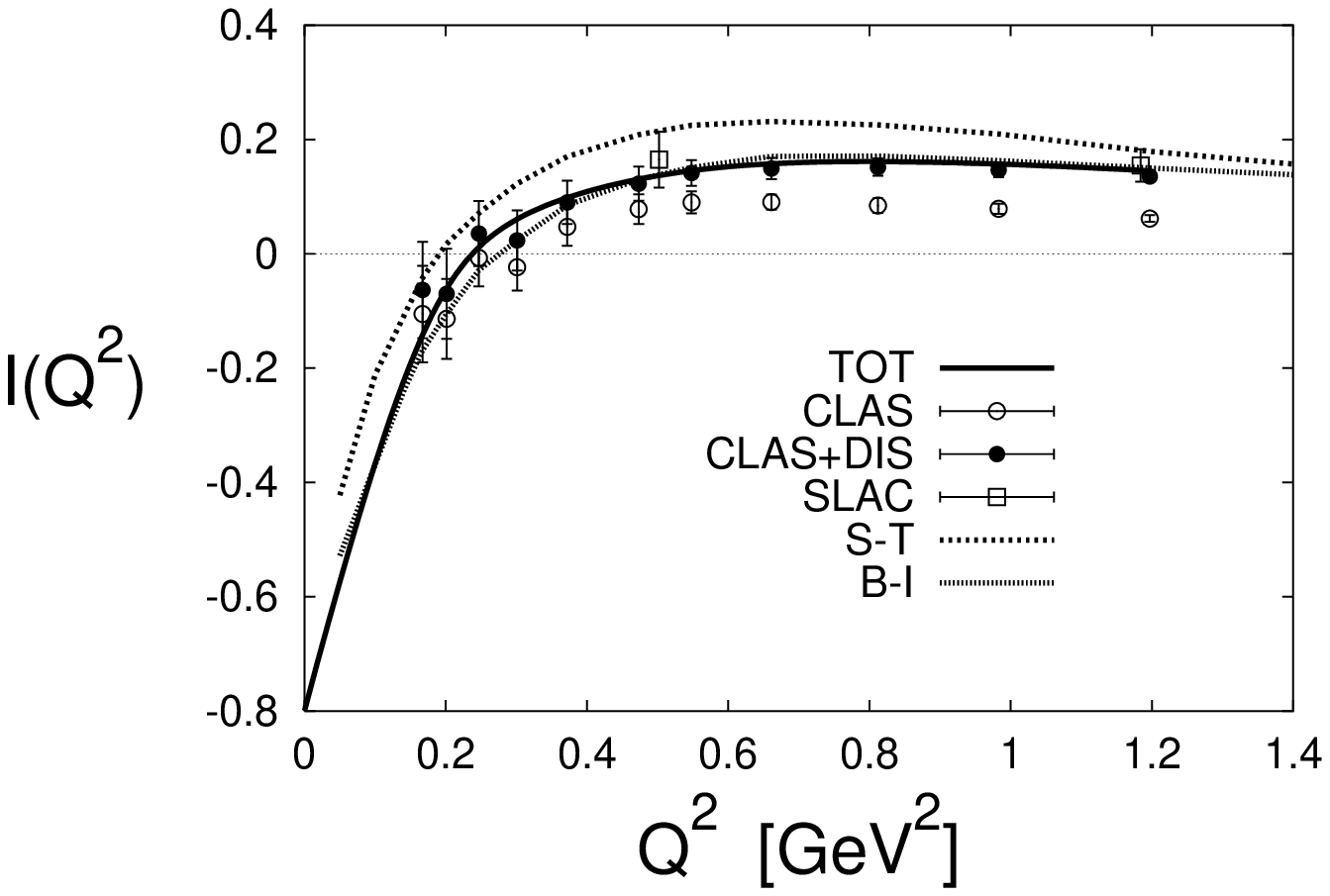}\\
\end{center}
\caption{The DHGHY moment $I(Q^2)$ for the proton with the VMD part parametrized
using the GRSV fit \cite{grsv2000}.
Shown are also calculations of \cite{burkert} 
(``B--I'') and \cite{soffer} (``S--T''). Points marked ``CLAS'' result from 
the JLAB E91-023 experiment using the Cebaf Large Angle Spectrometer, CLAS
\cite{e91-023}: the open circles refer to the resonance region, $W < W_t(Q^2)$, 
and the full circles contain a correction for the DIS contribution. 
Points marked ``SLAC'' are from \cite{nne143}. Errors are total.
}
\label{fig3}
\end{figure}
%
%
\section{Numerical calculations for the proton}

\noindent  
To obtain the value of $C$ from Eq. 
(\ref{i0}), the contribution of resonances was evaluated using the preliminary
 data taken at ELSA/MAMI by the GDH Collaboration \cite{elsa} 
at the photoproduction, for $W_t$=1.8 GeV.  
Asymptotic part of $g_1$  was parametrized using GRSV2000 fit
for the ``standard scenario'' of polarized parton distributions with
a flavour symmetric light sea, $\Delta\overline u = \Delta\overline d = 
\Delta s = \Delta\overline s$,  at the NLO accuracy \cite{grsv2000}. 
The nonperturbative parton distributions, $\Delta p_j^{(0)}(x)$ in the
light vector meson component of $g_1$, Eq.(\ref{vmd2}) were evaluated at 
fixed $Q^2 = Q_0^2$, using, either (i) the GRSV2000 fit, 
or (ii) a simple, ``flat'' input: 
%
%
%
%
%
\begin{equation}
\Delta p_i^{(0)}(x)=N_i (1-x)^{\eta_i}
\label{dpi0}
\end{equation}
with $\eta_{u_v}=\eta_{d_v}=3,$  $\eta_{\bar u} = \eta_{\bar s} = 7 $
and $\eta_g=5$. Normalization constants $N_i$ were determined
by imposing the Bjorken sum rule for $\Delta u_v^{(0)}-\Delta d_v^{(0)}$,
and requiring that the first moments of all other distributions are the
same as those determined from  the QCD analysis \cite{STRATMAN}.
It was checked that parametrization (\ref{dpi0}) combined with
the unified equations gives reasonable description of the SMC data
on $g_1^{NS}(x,Q^2)$ \cite{bbjk} and on $g_1^p(x,Q^2)$ \cite{bkk}. 
This fit was also
used to investigate the magnitude of the double logarithmic corrections,
ln$^2(1/x)$, to the spin structure function of proton at low $x$ \cite{kz}.
\noindent
We have assumed $Q_0^2$ = 1.2 GeV$^2$, cf. Eqs. (\ref{gvmd}) and (\ref{vmd2}), 
in accordance with the analysis of $F_2$, \cite{el89,el92}.
As a result the constant $C$ was found to be --0.30 in case (i) and
--0.24 in case (ii). These values change at most by 13$\%$ when  
$Q_0^2$ changes in the interval 1.0$ < Q_0^2 < $1.6 GeV$^2$.

\noindent
Negative value of the nonperturbative, Vector Meson Dominance, 
contribution was also obtained 
within a formalism based on unintegrated spin dependent parton distributions 
supplemented with the VMD \cite{bkk} and from the phenomenological analysis
of the sum rules \cite{ioffe2,burkert}.

\noindent
Using the above values of $C$, we have calculated \fsn from Eq. (\ref{gvmdu}).
Results are shown in Fig.\ref{fig0} for different values of $x$ and $Q^2$, 
separately for the VMD and asymptotic parts. The VMD decreases fast with $Q^2$;
for $Q^2\sim$ 10 GeV$^2$ the asymptotic function coincides to better
than 0.1 $\%$ with $g_1(x,Q^2)$.

\noindent
The only measurements performed at low values 
of $x$ and $Q^2$ are those obtained by the SMC with a dedicated low $x$ trigger,
\cite{ssmc98}. They are plotted in Fig.\ref{fig1} together with 
the results of
our model, calculated at the $(x,Q^2)$ values corresponding to those of the
SMC. Mean $Q^2$ at lowest (highest) $x$ is 0.02 GeV$^2$ (0.63 GeV$^2$).
 Kinematic region of that experiment corresponds to $x<x_t=Q^2/2M\nu_t(Q^2)$,
i.e. it lays outside the baryonic resonance region. Our model reproduces well 
a general trend in the data; however experimental errors are too large for 
a more detailed analysis.
 
\noindent
We have also computed the DHGHY moment, Eq.(\ref{iq2}), for the proton.
As a resonance input we used 
the preliminary results of the JLAB E91-023 experiment \cite{e91-023}
for 0.15$\lapproxeq Q^2 \lapproxeq $1.2 GeV$^2$ and $W < W_t=W_t(Q^2)$~\cite{fatemi}.
Results are shown in Fig.\ref{fig2}, 
separately for contributions from the VMD, from the 
asymptotic part and from the resonances.   
Partons contribute significantly even in the photoproduction limit where the
main part of the $I(Q^2)$ comes from resonances.

\noindent
In Fig. \ref{fig3} we show our DHGHY moment together with the results of
calculations of Refs \cite{burkert,soffer} as well as with
the SLAC and E91-023 measurements in the resonance region; the latter were
 used as an 
input to our $I(Q^2)$ calculations. We also show the E91-023 data 
corrected by their authors for the deep inelastic contribution. 
Our calculations are slightly 
larger than the DIS-corrected data and then the results of \cite{burkert} but
clearly lower than the results of \cite{soffer} which overshoot the data.

\section{Conclusions and outlook}

\noindent
We have analysed the spin dependent structure function \fsn at low values 
of $Q^2$ and $x$ in the
framework of the Generalised Vector Meson Dominance model. This model was very
successful in describing the behaviour of the unpolarized structure function
$F_2$ in the same kinematic region. Contributions from
both light- 
and heavy 
vector mesons have been evaluated. 
The latter part of the structure function is directly related to the 
(QCD improved) parton model $g_1$ and was evaluated using the GRSV2000 fit
in the standard scenario at the NLO accuracy. The light meson contribution
represents the nonperturbative part of $g_1$ and was parametrized either through
combinations of the GRSV2000 parton distributions evaluated at fixed 
$Q^2=Q_0^2$, or through combinations of (almost) $x-$independent
phenomenological  parton distributions.
Contribution from the nonperturbative part was  fixed using the
Drell-Hearn-Gerasimov-Hosoda-Yamamoto sum rule which is related to the
first moment of $g_1$ in the photoproduction limit. To employ that property 
an independent information about the resonance region was necessary. 
We used preliminary measurements of the GDH Collaboration at ELSA/MAMI and of
the JLAB E91-023 experiment. As a result the contribution
of the nonperturbative part of $g_1$ was found to be negative and equal to
about 0.24 - 0.30, depending on the parametrization. This result changes by
at most 13$\%$ for the $Q^2$ in the interval 1.0$ < Q^2 <$ 1.6 GeV$^2$.  
The final spin dependent structure
function \fsn  reproduces well the trends in the low $x$, low $Q^2$ data
of the SMC but for more detailed comparison more precise measurements are
needed. Hopefully they will
 soon be performed by the COMPASS experiment at CERN. We have also 
computed the DHGHY integral which reproduces well the DIS corrected
preliminary measurements by JLAB E91-023.

 
\section*{Acknowledgments} 
 
This research has been supported in part by the Polish Committee for Scientific  Research with grants 2 P03B 05119, 2PO3B 14420 and European 
Community grant 'Training and Mobility of Researchers', Network 'Quantum      
Chromodynamics and the Deep Structure of Elementary Particles'      
FMRX-CT98-0194. B.\ Z.\ was supported by the Wenner-Gren Foundations.
%

\end{document}